\RequirePackage{fixltx2e}
\documentclass[aps,prl,reprint,floatfix,nobibnotes,superscriptaddress]{revtex4-1}

\usepackage{graphicx}
\usepackage{dcolumn}
\usepackage{bm}
\usepackage{amsmath,amssymb}
\usepackage{color}

\begin{document}

\title{Correlation between Fragility and the Arrhenius Crossover Phenomenon in Metallic, Molecular, and Network Liquids}

\author{Abhishek \surname{Jaiswal}}
\affiliation{Department of Nuclear, Plasma and Radiological Engineering,University of Illinois at Urbana-Champaign, Urbana, Illinois 61801, USA}

\author{Takeshi \surname{Egami}}
\affiliation{Department of Materials Science and Engineering, University of Tennessee, Knoxville, Tennessee 37996, USA and Oak Ridge National Laboratory, Oak Ridge, Tennessee 37831, USA}

\author{K. F. \surname{Kelton}}
\affiliation{Department of Physics and Institute of Materials Science and Engineering, Washington University, St. Louis, Missouri 63130, USA}

\author{Kenneth S. \surname{Schweizer}}
\affiliation{Department of Materials Science and Engineering, University of Illinois at Urbana-Champaign, Urbana, Illinois 61801, USA}

\author{Yang \surname{Zhang}}
\thanks{Corresponding author. E-mail: zhyang@illinois.edu}
\affiliation{Department of Nuclear, Plasma and Radiological Engineering,University of Illinois at Urbana-Champaign, Urbana, Illinois 61801, USA}
\affiliation{Department of Materials Science and Engineering, University of Illinois at Urbana-Champaign, Urbana, Illinois 61801, USA}


\date{\today}

\begin{abstract}
We report the observation of a distinct correlation between the kinetic fragility index $m$ and the reduced Arrhenius crossover temperature $\theta_A = T_A/T_g$ in various glass-forming liquids, identifying three distinguishable groups. In particular, for 11 glass-forming metallic liquids, we universally observe a crossover in the mean diffusion coefficient from high-temperature Arrhenius to low-temperature super-Arrhenius behavior at approximately $\theta_A \approx 2$ which is in the stable liquid phases. In contrast, for fragile molecular liquids, this crossover occurs at much lower $\theta_A \approx 1.4$ and usually in their supercooled states. The $\theta_A$ values for strong network liquids spans a wide range higher than 2. Intriguingly, the high-temperature activation barrier $E_\infty$ is universally found to be $\sim 11\ k_B T_g$ and uncorrelated with the fragility or the reduced crossover temperature $\theta_A$ for metallic and molecular liquids. These observations provide a way to estimate the low-temperature glassy characteristics ($T_g$ and $m$) from the high-temperature liquid quantities ($E_\infty$ and $\theta_A$).

\end{abstract}


\maketitle


The fragility of a glass-forming liquid is a measure of how quickly its dynamics slows down upon cooling. It is usually quantified by the kinetic fragility index $m$, which is defined as the slope of the Angell plot of transport coefficients in logarithmic scale versus $T_g/T$ evaluated at the glass transition temperature $T_g$ \cite{Angell2000b}:
\begin{equation}
\label{eq:eqn1}
m = \left. \frac{\partial \log \eta(T)}{\partial (T_g/T)} \right|_{T=T_g} 
\end{equation}
A liquid that undergoes little change in the slope as a function of temperature is called as a kinetically ``strong'' system. Examples include many of the network liquids such as silica, soda lime glasses, etc. The other end of the spectrum, defined as ``fragile'', corresponds to systems that show significant increases in slope with cooling. Examples include many of the van der Waals molecular liquids, polymers, ionic liquids, etc. Notably, many of the glass-forming metallic liquids that are mediated by complex many-body metallic interactions span the intermediate fragility range \cite{Chen2011a,Qin2006}. This fact has inspired us to systematically compare the nature of slow dynamics of these metallic liquids with other glass-formers, namely, molecular liquids and network liquids. To date, an agreed quantitative understanding of fragility is still lacking \cite{Angell1995,Debenedetti2001}. 

The glass transition temperature $T_g$ and the fragility index $m$ are key parameters quantifying the low-temperature behavior of liquids. The Arrhenius crossover phenomenon occurring well above $T_g$ quantifies the high-temperature activated behavior of liquids \cite{Dyre1998,Tarjus2000,Roland2008,Schmidtke2012,Mirigian2014,Novikov2016} and has attracted much attention recently, especially in metallic liquids \cite{Iwashita2013a,Blodgett2014,Mauro2014b}. For instance, in many molecular liquids the Arrhenius crossover or glassy-dynamics-onset temperature $T_A$ marks the deviation of the transport coefficients or the relaxation time from the high-temperature Arrhenius dependence as well as the deviation of the intermediate scattering function from a simple exponential relaxation. This crossover is believed to indicate increasingly dynamically heterogeneous and cooperative motion when temperature is lowered below $T_A$ \cite{Ediger2000b,Richert2002}. Above $T_A$, particles move relatively independently without the need for a collective reorganization of their respective local environment due to the large mobility and phonon localization \cite{Iwashita2013a}. However, when the temperature is lowered, collective reorganization of particles (local topological excitations \cite{Iwashita2013a}, hopping \cite{Kob1995b}, etc.) over increasing length scales is needed to facilitate large amplitude irreversible motions in cold dense media. Such cooperative motion allows the system to overcome large free energy barriers and relax, resulting in highly activated dynamics \cite{Mirigian2014,Sastry1998a,Berthier2011c,Chandler2010,Stevenson2006,Donati2002,Schroder2000}. 
Note that it is $T_A$, and not $T_g$, that marks the onset of such cooperativity and dynamic heterogeneity. Similarly, in strong liquids, cooperativity and spatially heterogeneous dynamics have also been observed with emerging slow dynamics. However, the apparent Arrhenius behavior of transport properties remains largely unchanged down to low temperatures since large scale cooperative motions are less relevant due to the rather uniform topography of the energy landscape and influence of strong covalent bonds \cite{Vogel2004,Coslovich2009}. There have been limited studies on the nature of the Arrhenius crossover in glass-forming metallic liquids until very recently \cite{Blodgett2014,Mauro2014b}. Consequently, a systematic comparison between metallic liquids and molecular and network liquids regarding the Arrhenius crossover phenomenon and its relation with the fragility is still missing.

In this Letter, we explore the relation between the dynamic fragility and the Arrhenius crossover phenomenon in 11 metallic liquids, 56 molecular liquids, and 12 network liquids. Both $T_g$ and $m$ are low-temperature parameters, while the Arrhenius crossover characterized by $E_\infty$ and $T_A$ is a high-temperature phenomenon. It is not obvious that a connection between these two phenomena exists, nor it is clear how this transpires across the various classes of glass-formers. To this end, we have observed a direct correlation between fragility and the Arrhenius crossover in all studied liquids. Three distinct regions in a fragility $m$  versus the reduced Arrhenius crossover temperature $\theta_A$ plot have been established. Strong network liquids reveal the highest crossover temperature (relative to $T_g$) to cooperative dynamics followed by metallic liquids that have intermediate fragilities. Fragile molecular liquids are found to have the lowest $\theta_A$ that usually occurs in their supercooled states, unlike metallic liquids. Furthermore, we also find, intriguingly, that the high-temperature effective activation energy for transport $E_\infty$ in metallic and molecular liquids are surprisingly similar, roughly $\sim 11\ k_B T_g$.

\begin{figure}[tbp!]
\includegraphics[width=\linewidth]{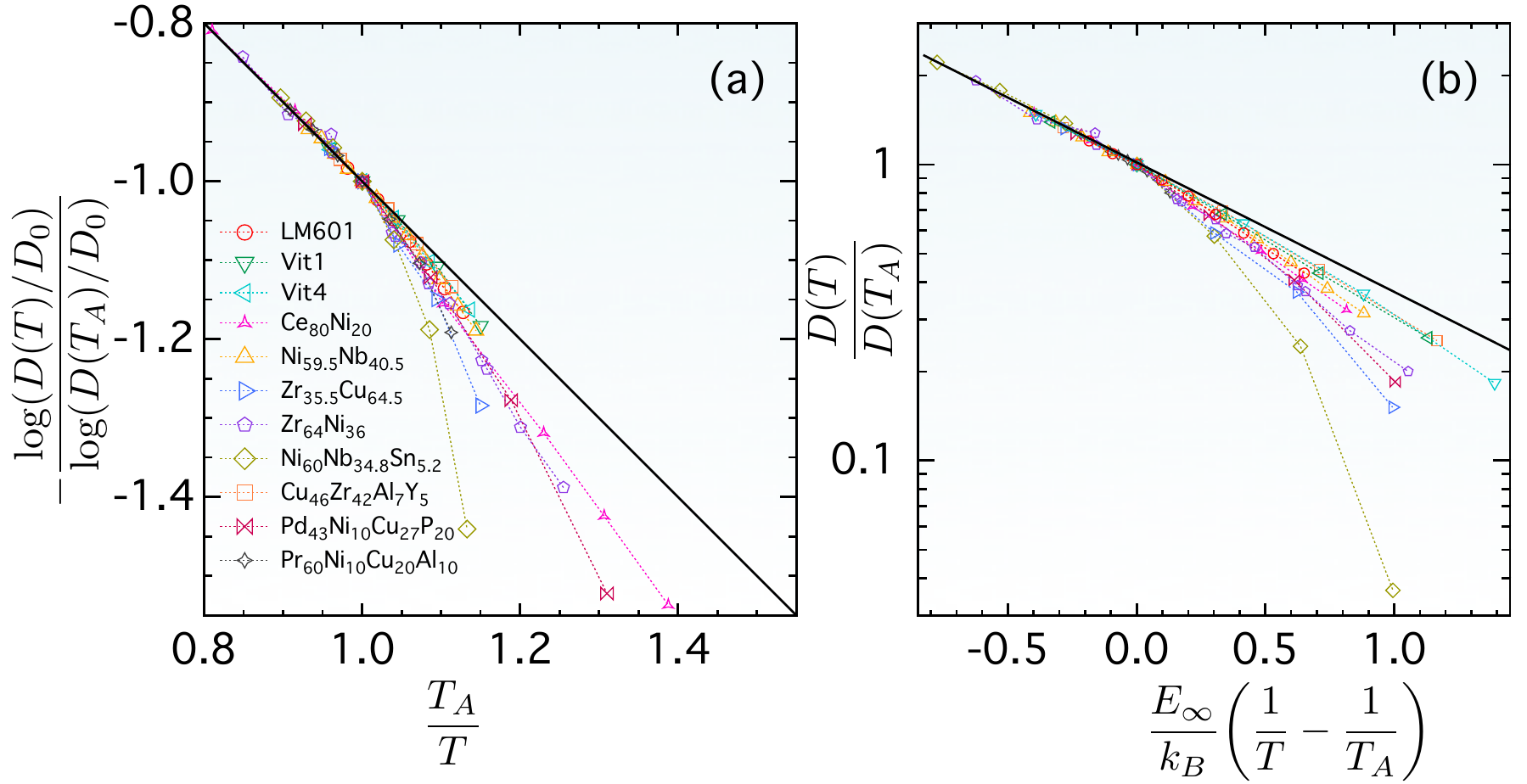}
\caption{\label{fig:diffCoeff} \textbf{Arrhenius crossover in 11 metallic liquids.} All samples presented here were measured using QENS. (a) and (b) show two different scaling plots of the mean diffusion coefficients. Solid lines represent the Arrhenius law. Both plots clearly show a deviation from the high-temperature Arrhenius law below $T_A$.}
\end{figure}

The onset of cooperative dynamics has been previously characterized in fragile liquids by studying the deviation of transport coefficients from their high-temperature Arrhenius behavior \cite{Mirigian2014,Schmidtke2012,Blodgett2014,Jaiswal2015a}. Herein, we compile diffusion coefficient data of 11 metallic liquids  \cite{Jaiswal2016,Meyer2002,Brillo2011,Chathoth2008a,Chathoth2008,Chathoth2009,Chathoth2010a,Chathoth2012,Yang2014,Yang2014d,Kordel2011b,Meyer2003b} measured using Quasi-Elastic Neutron Scattering (QENS) that measures the mean relaxation dynamics of multicomponent metallic liquids. Fig.~\ref{fig:diffCoeff} shows that all of the 11 glass-forming metallic melts exhibit a clear deviation from their high-temperature Arrhenius behaviors. It should be noted that relaxation in such metallic liquids are slightly stretched even in the very high temperature liquid state due to differences between the mobilities of constituent elements \cite{Jaiswal2016}. Nevertheless, previous studies has shown $T_A$ to be the same for all components \cite{Jaiswal2015a}. In Fig.~\ref{fig:diffCoeff}(a), we apply a straightforward scaling of the Arrhenius form of the diffusion coefficient:
\begin{equation}
\label{eq:eqn2}
\frac{D(T)}{D_0} = \exp\left(-\frac{E_\infty}{k_BT}\right) \Rightarrow \frac{\log(D(T)/D_0)}{\log(D(T_A)/D_0)} = \frac{T_A}{T}
\end{equation}
where $D_0$ is the diffusion constant and $E_\infty$ is the high-temperature activation barrier. Both $D_0$ and $E_\infty$ are obtained from fitting the high temperature data using the Arrhenius equation as shown on the left hand side of Eq.~(\ref{eq:eqn2}). The diffusion coefficients of all 11 metallic liquids collapse onto a single straight line with a slope of $-1$ above $T_A$ of the respective sample. Deviations from the straight line is unambiguously observed below $T_A$ for all the metallic liquids. An alternative scaling is also presented in Fig.~\ref{fig:diffCoeff}(b) where $D_0$ is divided out, as shown in Eq.~(\ref{eq:eqn3}). 
\begin{equation}
\label{eq:eqn3}
\frac{D(T)}{D(T_A)} = \exp \left( - \frac{E_\infty}{k_B} \left( \frac{1}{T} - \frac{1}{T_A} \right) \right)
\end{equation}
This scaling also reveals consistent Arrhenius crossover behavior as described above. The timescale associated with this crossover is typically around $10-20$~ps in metallic liquids, which is consistent with recent experimental observations \cite{Roland2008,Schmidtke2012}, and a prediction of the elastically collective activated hopping theory, of a crossover time of about $10-100$~ps for molecular liquids \cite{Mirigian2014}. For metallic liquids, this timescale is estimated assuming a Fickian behavior $\tau_A \sim d^2/(6D(T_A))$, where $d$ is the average particle diameter. $\tau_A$ is much larger than boson peak timescale in metallic liquids that typically occurs at $\sim$5 meV or $\sim$0.13~ps. In fact, the $Q$-dependent relaxation time measured by QENS spans a range of $1-100$~ps.  

\begin{figure}[htbp!]
\includegraphics[width=0.75\linewidth]{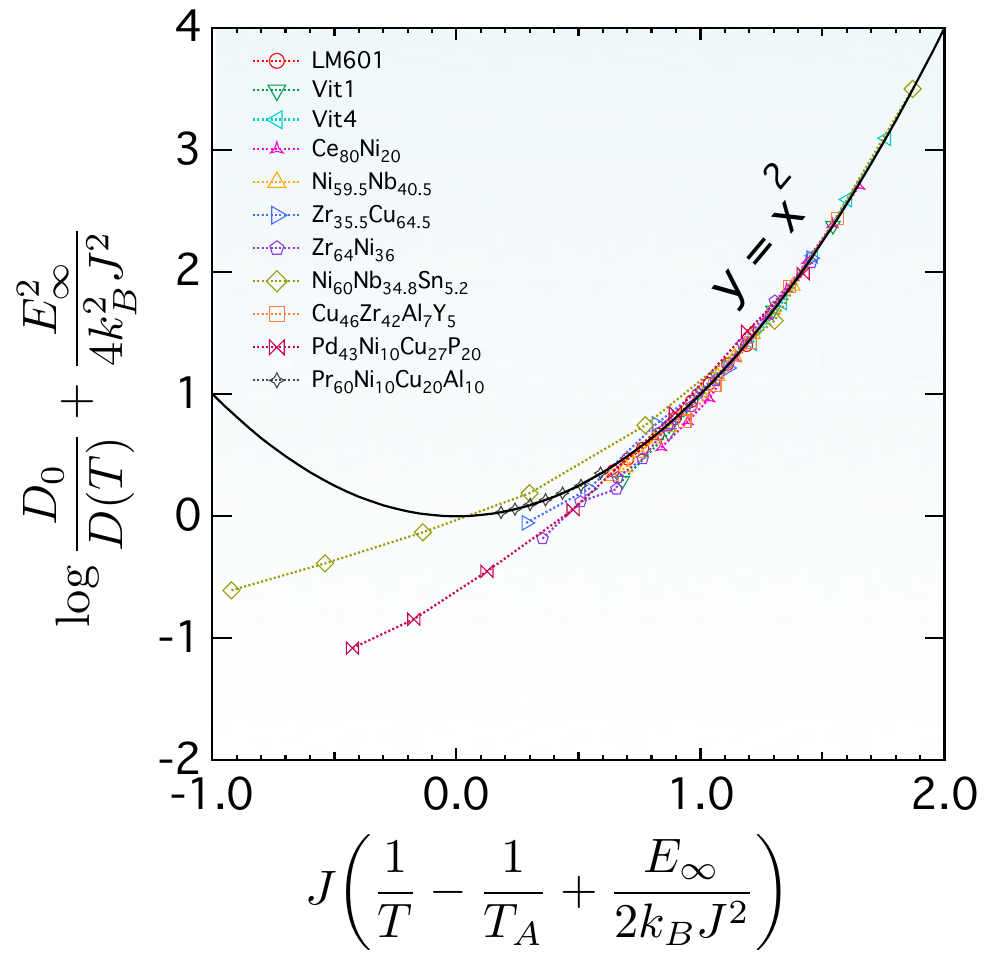}
\caption{\label{fig:LTscaling} \textbf{Scaling of the low-temperature diffusion coefficient of 11 metallic liquids using the parabolic formula of Eq.~(\ref{eq:eqn3}).}  Collapse of low-temperature diffusion coefficients to a single master parabolic curve is evident. A clear deviation from the parabolic form is observed above $T_A$ for a given metallic liquid.}
\end{figure}

Below the crossover temperature $T_A$, several analytical forms have been developed to model the super-Arrhenius dependence of transport coefficients. We chose the parabolic formula as a convenient analysis tool with the assumption that the onset temperature of dynamical facilitation \cite{Keys2011} is close to $T_A$ and is identified as deviations from the parabolic form, although our analysis does not prove the ``onset temperature'' is identical to $T_A$, nor that the low temperature physics is facilitation given that different activated theories predict the similar formula \cite{Mirigian2014}. The parabolic formula when applied to metallic liquids needs to be supplemented by an additional non-vanishing mean-field energy contribution:
\begin{equation}
\label{eq:eqn4}
\frac{D_0}{D(T)} = \exp\left(J^2\left( \frac{1}{T} - \frac{1}{T_A}\right)^2 +\frac{E_\infty}{k_B}\left(\frac{1}{T} - \frac{1}{T_A} \right) \right)
\end{equation}
where the activation energy $E_\infty$ and the crossover temperature $T_A$ are determined from the high-temperature Arrhenius fitting. A simple scaling (details in the SI \cite{SI}), shows that the inverse diffusion coefficient converges towards the parabolic form at low temperatures, as shown in Fig.~\ref{fig:LTscaling}. The point at which the experimental data deviates from the $y=x^2$ curve (solid black line) is the crossover temperature of a given metallic liquid. Such a low-temperature scaling plot verifies the consistency in determining $T_A$.

In our recent simulations \cite{Jaiswal2015a}, we found that the Arrhenius crossover is associated with a sudden increase in the size of dynamical clusters of particles, of notably slow to intermediate mobility. This occurs at roughly the same temperature for all constituent elements in metallic liquids. Due to the presence of temporal clusters of varying mobility there is increasing heterogeneous dynamics, which was further validated by quantitative measures such as the non-Gaussian parameter and the four-point correlation functions. Below $T_A$, the Stokes-Einstein relation begins to break down corresponding to a decoupling of the diffusive and relaxation dynamics for all components. It should be noted that the Arrhenius crossover discussed in this paper is not identical to other dynamical crossovers, such as the fragile-to-strong crossover, the empirically-deduced mode-coupling crossover, the separation of $\alpha$ and the Johari-Goldstein $\beta$ relaxation, and others, which typically occur at lower temperatures than $T_A$ \cite{Mallamace2010,Zhou2015,Novikov2003b,Rossler1998,Schweizer2004,Yu2013z,Yu2014a}.

\begin{figure}[tbp!]
\includegraphics[width=\linewidth]{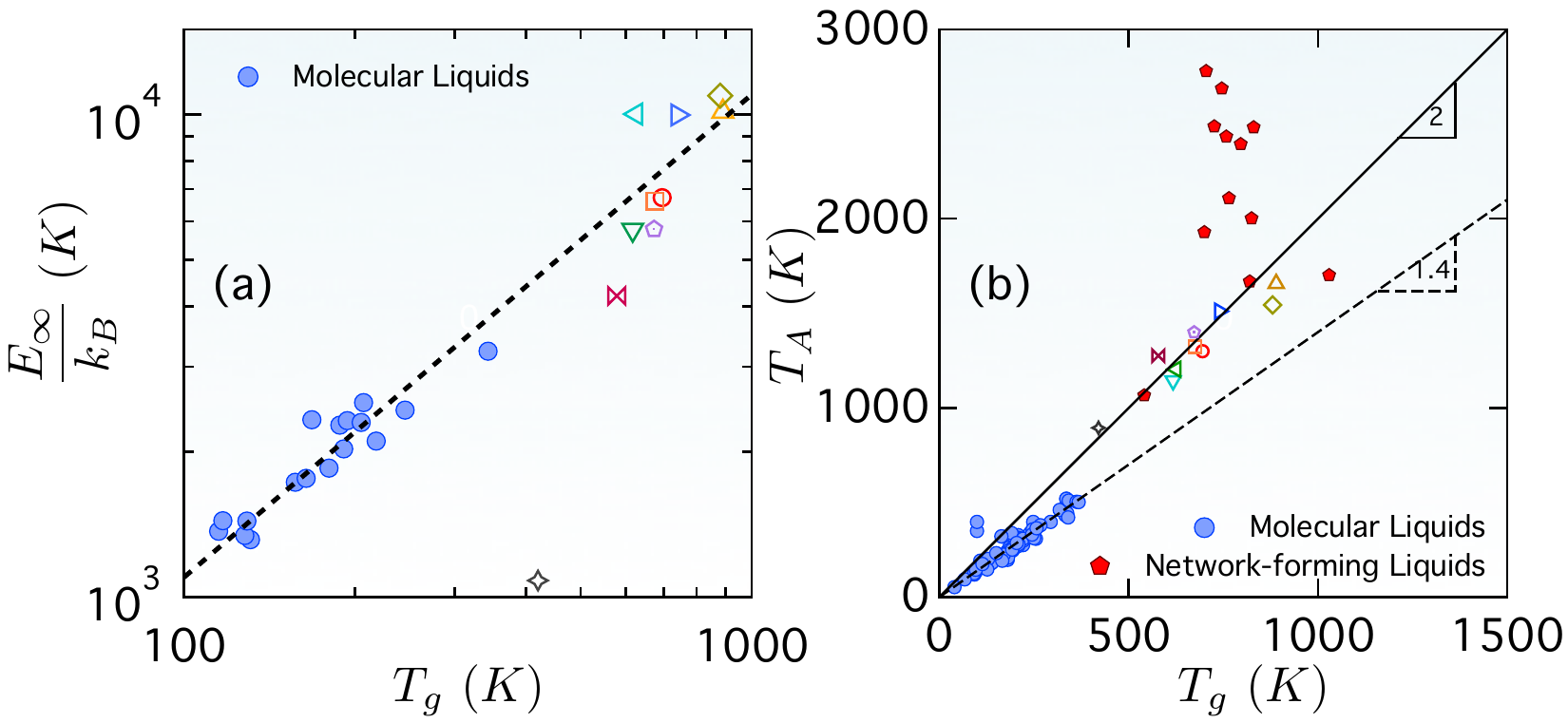}
\caption{\label{fig:ActEnergy} \textbf{Correlations of the high-temperature activation energy barrier $E_\infty$ and the Arrhenius crossover temperature $T_A$ with the glass transition temperature $T_g$.} (a) In all metallic and molecular liquids, $E_\infty$ is universally $\sim 11\ k_B T_g$, independent of their fragilities. While for strong network liquids, $E_\infty \approx E(T_g) = m\ k_B T_g$. (b) For 11 metallic liquids, $T_A \approx 2\ T_g$. For 56 molecular liquids, $T_A \approx 1.4\ T_g$. For 12 network liquids, there is no clear correlation of $T_A$ with $T_g$, which is expected.}
\end{figure} 

Once the crossover temperature is identified, we can examine $E_\infty$ and $T_A$ for a variety of liquids with diverse fragilities. The effective activation barrier $E_\infty$ for the high-temperature Arrhenius diffusion process in all metallic liquids is found universally to be $\sim 11\ k_B T_g$, as shown in Fig.~\ref{fig:ActEnergy}(a). The notable outliers are Vitreloy 4 (left facing triangle) and the Pr-based glass-former (diamond). In both cases, the reason for such variations in $E_\infty$ is likely due to the very limited temperature range of the diffusion coefficient measurements. The activation barrier for the viscosity of Vitreloy 4 is 55~kJ/mol \cite{Basuki2014a}, which is very similar to that for other Cu-Zr based systems close to $11\ k_B T_g$. Interestingly, such an activation barrier of $E_\infty \sim 11\ k_B T_g$ is surprisingly similar to that of many van der Waals molecular liquids, as recently established experimentally \cite{Petzold2013,Schmidtke2012,Kivelson1996} and also predicted by the microscopic elastically collective activated dynamics theory \cite{Mirigian2014} (see SI for details). 
This behavior is independent of the fragility of these two classes of liquids, as discussed in the SI. Note that the hydrogen bonded systems and long chain polymers are characterized by a higher $E_\infty$ \cite{Mirigian2014,Schmidtke2015}. For strong network liquids, $E_\infty \approx E(T_g) = m\ k_B T_g$ because of weak changes in the slope of transport properties in the Angell plot. Consequently, their reduced activation energy $E_\infty/k_B T_g$ is as large as their fragility index $m$, typically in the range of 20 -- 30. For example, for many silicate and borosilicate based liquids, $E_\infty/k_B T_g$ is indeed very close to their reported fragility index $m$ \cite{Novikov2004}. 

Perhaps the more interesting result is obtained by comparing $T_A$ with $T_g$ for the three classes of glass-formers as shown in Fig.~\ref{fig:ActEnergy}(b). Three distinct behaviors can be identified: 1) For metallic liquids, it is remarkable that all data points follow a straight line with a slope of 2. In our previous QENS experiments, we have made a similar observation of $T_A \approx 2\ T_g$ for diffusion in LM601 
\cite{Jaiswal2016,Blodgett2014}. Here we find such a relation is universal in all metallic glass-formers examined, which are composed of two to five elements. Furthermore, $T_A$ is found to be higher than the melting temperature $T_m$, and thus the dynamic crossover occurs in the equilibrium liquid state. These observations of an Arrhenius crossover in the diffusion coefficient are in excellent agreement with recent results from studies of the shear viscosity in 27 glass-forming metallic liquids \cite{Blodgett2014,Mauro2014b}. The $T_{A}$ identified from deviation of the bulk viscosity from an Arrhenius behavior at high temperatures was also found to be $\sim$2~$T_g$. It should be noted that the crossover observed in macroscopic viscosity (associated with collective stresses) does not a priori imply that an Arrhenius crossover will occur in a microscopic diffusion process at the same place because metallic liquids have been found to violate the Stokes-Einstein relation even above the melting point and in the vicinity of $T_A$ \cite{Brillo2011,Chathoth2010a}. Our new results suggest a physical picture of the dynamic crossover in metallic melts in terms of the single particle self-diffusion coefficient, which is qualitatively consistent with reported results on viscosity. 
\begin{figure}[tbp!]
\includegraphics[width=0.75\linewidth]{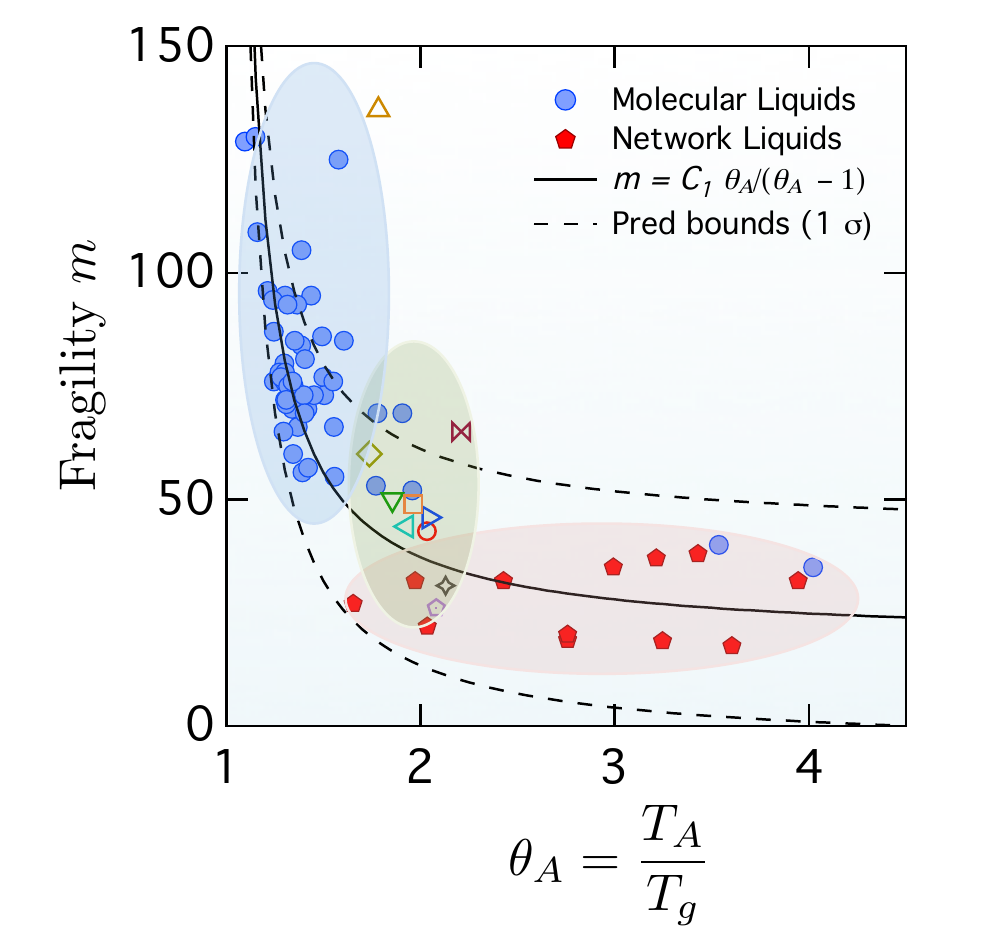}
\caption{\label{fig:Onset} \textbf{Correlations of the fragility index $m$ with the reduced Arrhenius crossover temperature $\theta_A$ for various glass-formers.} An inverse relation is observed between the two quantities. The analytical expression (Eq.~(\ref{eq:eqn5})) is applied to the data.}
\end{figure}
2) For molecular liquids, the crossover temperatures $T_A$ are typically found \cite{Elmatad2009}, and recently predicted theoretically \cite{Mirigian2014}, to be $1.4\pm0.2\ T_g$. As seen in Fig.~\ref{fig:ActEnergy}(b), almost all the data points fall on the line with slope of 1.4. The two clear outliers represent two different data sets for propanol, a molecule that can form hydrogen bonds. Similarly, other hydrogen bonded systems such as glycerol and sorbitol also display higher relative crossover temperatures of $T_A \sim 1.55 - 1.75\ T_g$ \cite{Mirigian2014}. 
3) For network liquids, the crossover temperatures $T_A$ span a very wide range and are typically higher than $2~T_g$. This is because many of these systems are kinetically strong and thus do not show a distinct deviation from the high-temperature Arrhenius behavior. Consequently, there are large uncertainties in identifying the crossover temperature. 

In Fig.~\ref{fig:Onset}, we plot the the fragility index $m$ versus the reduced crossover temperature $\theta_A = T_A/T_g$. One can readily see there are three distinct regions associated with the chemically different nature of the glass-formers. At a rough qualitative level, $m$ is found to be inversely proportional to $\theta_A$. The highly fragile molecular liquids show a much lower $\theta_A$; metallic liquids have a higher $\theta_A$, corresponding to their intermediate fragilities; while the network liquids have very small fragilities but cover a wide range of $\theta_A$. The separation into different groups is likely due to the proximity of the fragilities of liquids in each group. We anticipate other important class of glass-formers such as ionic liquids ($m$ in the range of 50 -- 100 \cite{Qin2006}, only CaKNO$_3$ shown here) and chalcogenides ($m$ in the range of 30 -- 70 \cite{Svoboda2015} similar to metallic liquids, not discussed in this work) should likely fill in the gaps in between and around $\theta_A \sim 1.4$ and $\theta_A \sim 2$ in Fig.~\ref{fig:Onset} respectively. In fact, a simple relation between $m$ and $\theta_A$ can be established from the parabolic equation \cite{Chandler2010, Keys2011}, which connects the onset behavior of glassy dynamics at $T_A$ to the temperature dependence of transport properties. Evaluating Eq.~(\ref{eq:eqn1}) yields the following relation (details in SI):
\begin{equation}
\label{eq:eqn5}
m = \frac{2J^2}{T_g}\left(\frac{1}{T_g} - \frac{1}{T_A}\right) = C_1 \frac{\theta_A}{\theta_A - 1} 
\end{equation}
where the prefactor $C_1 = \log(\tau(T_g)/\tau_0)$. We observe that this equation fits the reduced Arrhenius temperature dependence of the fragility admirably in Fig.~\ref{fig:Onset}(a). Since the parabolic form is valid only below the crossover temperature $T_A$, $\tau_0$ cannot be set not equal to the typical inverse phonon frequency value of $10^{-14}$ s. From our fittings we obtain $C_1 \approx 18$. The prediction bounds narrows for molecular liquids whose fragility values have been rigorously established giving rise to an almost constant $\theta_A \approx 1.4T_g$. For metallic liquids, the range of fragility values is very narrow hence giving rise to an almost constant $\theta_A$ around 2. The large uncertainty associated with $\theta_A$ for network liquids yields a wide prediction bound for fragility values.

We speculate that the distinction of $\theta_A$ among metallic, molecular, and network liquids can be rationalized, at least partially, in terms of the degree of harmonicity and softness of the pair interaction potential. The interatomic potential of metallic liquids is influenced by Friedel oscillations, and is harmonic and thus soft near its minimum. These features of the potential lead to low fragility \cite{Krausser2015}, similar to tunably soft cross-linked microgel colloids where both experiment \cite{Mattsson2009} and theory \cite{Yang2011} find softer repulsions correspond to lower dynamic fragility. For the same reason, the characteristic local atomic structure and connectivity persist at high temperatures, giving rise to a higher reduced crossover temperature $\theta_A$ in metallic liquids than typical van der Waals liquids, even higher than some hydrogen bonded molecular liquids such as glycerol and sorbitol. The network liquids are characterized by strong and directional covalent bonds \cite{Sidebottom2015}. The latter feature implies relaxation can be achieved by spatially local ``bond-breaking'' events, with a well defined activation energy, which results in very low fragilities and very high $\theta_A$. On the other end of the spectrum is molecular liquids which have strongly anharmonic interactions characterized by steep short-range repulsions. Therefore, their packing structures respond more sensitively to changes of density and temperature (a more ``fragile structure''), and collective molecular rearrangements beyond the first coordination shell only occur at much lower $\theta_A$. Furthermore, the fragility index $m$ has been shown to correlate with the elastic properties of glass-forming liquids such as the Poisson's ratio within a class of metallic or non-metallic glass-formers (but not universally across classes) \cite{Novikov2004,Novikov2005,Novikov2006,Johari2006}, composition \cite{Cheng2008k}, and elastic constants \cite{Wang2006}. Our results extend these correlations to the dynamics of the high-temperature liquid state.  

In summary, a clear correlation was found between the dynamic fragility and the Arrhenius crossover phenomenon. The reduced crossover temperature $\theta_A$ depends strongly on the liquid fragility, and can be observed either in the supercooled state (molecular glass-former) or in the equilibrium liquid state (metallic and network glass-formers). The effective activation barrier of the high-temperature Arrhenius behavior takes on a nearly universal value of $11\ k_B T_g$ for nonpolar molecular and metallic liquids. Such correlations between the low and high-temperature parameters imply that $T_g$ can be estimated from the high-temperature activation barrier $E_\infty$ and the fragility $m$ can be estimated from the reduced crossover temperature $\theta_A = T_A/T_g$. Hence, the low-temperature glassy characteristics can be predicted from the high-temperature Arrhenius crossover in liquids.


This work is supported by the U.S. Department of Energy, Office of Science, Office of Basic Energy Sciences, Materials Sciences and Engineering Division, under Award Number DE-SC-0014804.

\bibliography{all_refs}

\end{document}